\newcommand{\typeof}{0}
\documentclass[10pt,draft]{article}

\usepackage{ifthen}
\newcommand{\condinc}[2]{\ifthenelse{\equal{\typeof}{0}}{#1}{#2}}

\usepackage{proof}
\usepackage{amsmath}
\usepackage{amssymb}
\usepackage{subfigure}
\usepackage{epsf}
\usepackage{graphics}
\usepackage{wrapfig}
\usepackage{mathrsfs}
\usepackage{url}
\condinc{\usepackage{a4wide}}{}
\usepackage{calc}
\usepackage{amsmath,amssymb,amsfonts,mathrsfs,latexsym,stmaryrd}

\newcommand{\N}{\mathbb{N}}

\newcommand{\M}{\mathcal{M}}

\newcommand{\app}[2]{\{#1\}(#2)}
\newcommand{\timef}[1]{\mathit{Time}(#1)}

\newcommand{\timeapp}[2]{\mathit{Time}(\app{#1}{#2})}

\newcommand{\timearrow}[1]{\stackrel{#1}{\twoheadrightarrow}}
\newcommand{\timearrownp}{\twoheadrightarrow}
\newcommand{\cod}[2]{\ulcorner #1\urcorner^{#2}}
\newcommand{\btr}{\!\blacktriangleright\!}
\newcommand{\length}[1]{|#1|}
\newcommand{\truelength}[1]{||#1||}
\newcommand{\init}[2]{I(#1,#2)}
\newcommand{\final}[2]{F(#1,#2)}
\newcommand{\trans}[1]{T(#1)}
\newcommand{\function}[2]{U(#1,#2)}

\newcommand{\appendchar}[1]{\mathit{AC}(#1)}
\newcommand{\appendstring}[1]{\mathit{AS}(#1)}
\newcommand{\appendreverse}[1]{\mathit{AR}(#1)}
\newcommand{\convertchar}[2]{\mathit{CC}(#1,#2)}
\newcommand{\convertstring}[2]{\mathit{CS}(#1,#2)}
\newcommand{\Current}{\mathit{Current}}
\newcommand{\Preredex}{\mathit{Preredex}}
\newcommand{\Postredex}{\mathit{Postredex}}
\newcommand{\Function}{\mathit{Functional}}
\newcommand{\Argument}{\mathit{Argument}}
\newcommand{\Reduct}{\mathit{Reduct}}
\newcommand{\StackRedex}{\mathit{StackRedex}}
\newcommand{\StackTerm}{\mathit{StackTerm}}
\newcommand{\Counter}{\mathit{Counter}}

\condinc{
\newenvironment{varitemize}
{
\begin{list}{\labelitemi}
{\setlength{\itemsep}{0.0mm}
 \setlength{\topsep}{0.0mm}
 \setlength{\parindent}{0.0mm}
 \setlength{\parskip}{0.0mm}
 \setlength{\parsep}{0.0mm}
 \setlength{\partopsep}{0.0mm}
 \setlength{\leftmargin}{15pt}
 \setlength{\labelsep}{5pt}
 \setlength{\labelwidth}{10pt}}}
{
 \end{list} 
}}
{\newenvironment{varitemize}
{
\begin{list}{\labelitemii}
{\setlength{\itemsep}{0.0mm}
 \setlength{\topsep}{0.0mm}
 \setlength{\parindent}{0.0mm}
 \setlength{\parskip}{0.0mm}
 \setlength{\parsep}{0.0mm}
 \setlength{\partopsep}{0.0mm}
 \setlength{\leftmargin}{15pt}
 \setlength{\labelsep}{5pt}
 \setlength{\labelwidth}{10pt}}}
{
 \end{list} 
}}

\newcounter{number}

\newenvironment{numlist}
{\begin{list}{\arabic{number}.}
  {
   \usecounter{number}
   \setlength{\labelwidth}{4.0mm}
   \setlength{\labelsep}{2.0mm}
   \setlength{\itemindent}{0.0mm}
   \setlength{\itemsep}{0.0mm}
   \setlength{\topsep}{0.0mm}
   \setlength{\parskip}{0.0mm}
   \setlength{\parsep}{0.0mm}
   \setlength{\partopsep}{0.0mm}
  }
}
{\end{list}}

\title{\bf An Invariant Cost Model\\ for the Lambda Calculus}
\date{}
\condinc{
\author{{Ugo Dal Lago\footnote{
Dipartimento di Scienze dell'Informazione, Universit\`a di Bologna, 
Mura Anteo Zamboni 7, 40127 Bologna, Italy.
\texttt{dallago@cs.unibo.it}
}
}
\and 
{Simone Martini\footnote{
Dipartimento di Scienze dell'Informazione, Universit\`a di Bologna,
Mura Anteo Zamboni 7, 40127 Bologna, Italy.
\texttt{martini@cs.unibo.it}
}
}}}
{
\author{Ugo Dal Lago\and Simone Martini}
\institute{{Dipartimento di Scienze dell'Informazione, Universit\`a di Bologna\\
            via Mura Anteo Zamboni 7, 40127 Bologna, Italy\\
            \email{\{dallago,martini\}@cs.unibo.it}}}
}

\condinc{
\newtheorem{lemma}{Lemma}
\newtheorem{proposition}{Proposition}
\newtheorem{theorem}{Theorem}
\newtheorem{corollary}{Corollary}
\newenvironment{proof}{\begin{trivlist}
       \item[\hskip \labelsep {\bfseries Proof.}]}{\hfill $\Box$ \end{trivlist}}
\newtheorem{definition}{Definition}}{}

\begin{document}
\maketitle
\begin{abstract}
We define a new cost model for the call-by-value lambda-calculus satisfying
the invariance thesis. That is, under the proposed cost model, 
Turing machines and the call-by-value lambda-calculus can simulate each other 
within a polynomial time overhead. The model only relies
on combinatorial properties of usual beta-reduction, without
any reference to a specific machine or evaluator. In particular, the cost
of a single beta reduction is proportional to the difference between the size
of the redex and the size of the reduct. In this way, the total cost of normalizing a lambda term
will take into account the size of all intermediate results (as well
as the number of steps to normal form).
\end{abstract}
\section{Introduction}
Any computer science student knows that all computational models 
are extensionally equivalent, each of them characterizing the same class of computable functions.
However, the definition of \emph{complexity classes} by means of computational models must take into
account several differences between these models, in order to rule out unrealistic
assumptions about the cost of respective computation steps. It is then usual to consider only 
\emph{reasonable} models, in such a way that the definition of complexity classes
remain invariant when given with reference to any such reasonable model. 
If polynomial time is the main concern, this reasonableness requirement take the form
of the \emph{invariance thesis}~\cite{vanEmdeBoas90}:
\begin{quote}
\emph{Reasonable machines can simulate each other within 
a polynomially-bounded overhead in time and a constant-factor 
overhead in space}.
\end{quote}
Once we agree that Turing machines are reasonable, then many other machines
satisfy the invariance thesis. Preliminary to the proof of polynomiality of 
the simulation on a given machine, is the definition of a \emph{cost model}, 
stipulating when and how much one should account
for time and/or space during the computation. 
For some machines (e.g., Turing machines) this cost model is obvious; for others it is much less so.
An example of the latter kind is the type-free lambda-calculus, where there
is not a clear notion of \emph{constant time} computational step, and it is even less clear 
how one should count for consumed space.


The idea of counting the number of beta-reductions~\cite{DezaniRAIRO} is just too
na\"ive, because  beta-reduction is inherently 
too complex to be considered as an atomic operation,
at least if we stick to explicit representations of 
lambda terms. Indeed, in a beta step 
$$
(\lambda x.M)N\rightarrow M\{x/N\},
$$
there can be as many as $|M|$ occurrences
of $x$ inside $M$.
As a consequence, $M\{x/N\}$ can be as big as $|M||N|$.
As an example, consider the term $\underline{n}\;\underline{2}$,
where $\underline{n} \equiv \lambda x.\lambda y.x^n y$
is the Church numeral for $n$. Under innermost reduction
this term reduces to normal form in $3n-1$ beta steps, but there is an
exponential gap between this quantity and the time needed to write
the normal form, that is $\underline{2^n}$. Under outermost reduction, however,
the normal form is reached in an exponential number of beta steps. 
This simple example shows that taking the number of beta steps
to normal form as the cost of normalization is at least
problematic.  Which strategy should we choose\footnote{Observe 
that we cannot take the lenght of the longest reduction sequence, both because
in several cases this would involve too much useless work, and because for some
normalizing term there is \emph{not} a longest reduction sequence.
}? How do we
account for the size of intermediate (and final) results?\par


Clearly, a viable option consists in defining the cost of reduction
as
the time needed to normalize a term by another reasonable abstract machine, e.g. a
Turing machine. 
However, in this way we cannot compute the cost of reduction
from the structure of the term, and, as a result, it is very difficult to
compute the cost of normalization for particular
terms or for classes of terms.
Another invariant cost model is given by the actual cost of outermost (normal order) evaluation,
naively implemented~\cite{Lawall96icfp}. Despite its invariance, it is a too generous cost model
(and in its essence not much different from the one that counts the numbers of steps needed
to normalize a term on a Turing machine).
What is needed is a machine-independent, \emph{parsimonious}, and invariant
cost model. Despite some attempts~\cite{Frandsen91fpca,Lawall96icfp,Lawall97icfp} (which we will
discuss shortly), 
a cost model of this kind has not appeared yet.\par
To simplify things, we attack in this paper the problem for
the call-by-value lambda-calculus, where we do not reduce under an
abstraction and we always fully evaluate an argument before firing
a beta redex. Although simple, it is a calculus of paramount importance,
since it is the reduction model of any call-by-value functional
programming language. For this calculus we  define
a new, machine-independent cost model and we prove that it
satisfies the invariance thesis for time.  
The proposed cost model only relies
on combinatorial properties of usual beta-reduction, without
any reference to a specific machine or evaluator. The basic idea 
is to let the cost of performing a beta-reduction
step depend on the size of the involved terms. In particular, the cost
of $M\rightarrow N$ will be related to the \emph{difference}
$|N|-|M|$. In this way, the total cost of normalizing a lambda term
will take into account the size of all intermediate results (as well
as the number of steps to normal form). The last section of the paper will
apply this cost model to the combinatory algebra of closed lambda-terms, to establish
some results needed in~\cite{dallago05fsttcs}.
We remark that in this algebra the universal function (which maps two terms $M$ and $N$
to the normal form of $MN$) adds only a \emph{constant}
overhead to the time needed to normalize $MN$. This result, which is almost
obvious when viewed from the perspective of lambda-calculus, is something that cannot
be obtained in the realm of Turing machines.

\subsection{Previous Work}
The two main attempts to define a parsimonious cost model
share the reference to optimal lambda reduction \`a la L\'evy~\cite{Levy78phd}, a parallel strategy
minimizing the number of (parallel) beta steps (see~\cite{AG98}).

Frandsen and Sturtivant~\cite{Frandsen91fpca} propose a
cost model essentially based on the number of parallel beta
steps to normal form. Their aim is to propose a measure of efficiency
for functional programming language implementations. They show how
to simulate Turing machines in the lambda calculus
with a polynomial overhead. However, the paper does not 
present any evidence on the existence of a polynomial simulation in the other direction.
As a consequence, it is not known whether their proposal is invariant.\par
More interesting contributions come from the literature of the
nineties on optimal lambda reduction. Lamping~\cite{Lamping90popl}
was the first to operationally present this strategy as a graph rewriting
procedure. The interest of this technique for our problem stems
from the fact that a single beta step is decomposed into several
elementary steps, allowing for the duplication
of the argument, the computation of the levels of nesting
inside abstractions, and additional bookkeeping work. 
Since any such elementary step is realizable
on a conventional machine in constant time, Lamping's algorithm provides a theoretical
basis for the study of complexity of a single beta step.
Lawall and Mairson~\cite{Lawall96icfp} give results
on the efficiency of optimal reduction algorithms, highlighting the
so-called bookkeeping to be the bottleneck from the point ot view
of complexity. A consequence of Lawall and Mairson's work is evidence
on the inadequacy of the cost models proposed by Frandsen and 
Sturtivant and by Asperti~\cite{Asperti96popl},
at least from the point of view of the invariance thesis. In
subsequent work~\cite{Lawall97icfp}, Lawall and Mairson proposed a
cost model for the lambda calculus based on L\'evy's labels. They
further proved that Lamping's \emph{abstract} algorithm satisfies the proposed
cost model. This, however, does not imply by itself the existence of
an algorithm normalizing \emph{any} lambda term with a polynomial
overhead (on the proposed cost). Moreover, studying the dynamic
behaviour of L\'evy labels is clearly more difficult than dealing
directly with the number of beta-reduction steps. 

\section{Syntax}
The language we study is the pure untyped
lambda calculus endowed with lazy evaluation (that is, we never reduce
under an abstraction) and
call-by-value reduction. 
\begin{definition}
The following definitions are standard:
\begin{varitemize}
  \item
    Terms are defined as follows:
    $$
    M::=x\;|\;\lambda x.M\;|\;MM
    $$
    $\Lambda$ denotes the set of all lambda terms.
  \item
    Values are defined as follows:
    $$
    V::=x\;|\;\lambda x.M
    $$
    $\Xi$ denotes the set of all closed values.
  \item
    Call-by-value reduction is denoted by $\rightarrow$
    and is obtained by closing the rule
    $$
    (\lambda x.M)V\rightarrow M\{V/x\}
    $$
    under all applicative contexts. Here $M$ ranges
    over terms, while $V$ ranges over values.
  \item
    The length $\length{M}$ of $M$ is
    the number of symbols in $M$.
\end{varitemize}
\end{definition}
Following~\cite{PaoliniRonchi04} we consider this system as a complete calculus
and not as a mere strategy for the usual lambda-calculus. Indeed,
respective sets of normal forms are different. Moreover,
the relation $\rightarrow$ is not deterministic although, 
as we are going to see, this non-determinism is completely harmless.\par
The way we have defined beta-reduction implies a strong correspondence
between values and closed normal forms:
\begin{lemma}
Every value is a normal form and every 
closed normal form is a value.
\end{lemma}
\condinc{
\begin{proof}
By definition, every value is a normal form, because
evaluation is lazy and, as a consequence, every abstraction
is a normal form. For the other direction, we have to prove that
if $M$ is a closed normal form, then $M$ is an abstraction.
We proceed by induction on $M$. But if $M$ is an application $NL$,
then by induction hypothesis both $N$ and $L$ are 
abstractions and $M$ is not a normal form.
\end{proof}}{}
The prohibition to reduce under abstraction enforces a
strong notion of confluence, the so-called one-step diamond property,
which instead fails in the usual lambda calculus.
\begin{proposition}[Diamond Property]\label{prop:diamprop}
If $M\rightarrow N$ and $M\rightarrow L$ then
either $N\equiv L$ or there is $P$ such that 
$N\rightarrow P$ and $L\rightarrow P$.
\end{proposition}
\condinc{
\begin{proof}
By induction on the structure of
$M$. Clearly, $M$ cannot be a variable 
nor an abstraction so $M\equiv QR$.
We can distinguish five cases:
\begin{varitemize}
  \item
  If $Q\equiv\lambda x.T$ and $R$ is a
  value, then $N\equiv L\equiv T\{x/R\}$,
  because $R$ is a variable or an
  abstraction.
  \item
  If
  $N\equiv TR$ and
  $L\equiv UR$, where
  $Q\rightarrow T$ and
  $Q\rightarrow U$, then we 
  can apply the induction hypothesis.
  \item
  Similarly, if $R\rightarrow T$ and
  $R\rightarrow U$, where
  $N\equiv QT$ and
  $L\equiv QU$, then we 
  can apply the induction hypothesis.
  \item
  If $N\equiv QT$ and
  $L\equiv UR$, where 
  $R\rightarrow T$ and
  $Q\rightarrow U$, then
  $N\rightarrow UT$ and
  $L\rightarrow UT$.
  \item
  Similarly, if $N\equiv UR$ and
  $L\equiv QT$, where 
  $R\rightarrow T$ and
  $Q\rightarrow U$, then
  $N\rightarrow UT$ and
  $L\rightarrow UT$.  
\end{varitemize}
This concludes the proof.
\end{proof}}{}
As an easy corollary of Proposition~\ref{prop:diamprop}
we get an equivalence between all normalization strategies--- once again
a property which does not hold in the ordinary lambda-calculus.
\begin{corollary}[Strategy Equivalence]
$M$ has a normal form iff $M$ is strongly normalizing.
\end{corollary}
\condinc{
\begin{proof}
Observe that, by Proposition~\ref{prop:diamprop},
if $M$ is diverging and $M\rightarrow N$,
then $N$ is diverging, too. Indeed, if 
$M\equiv M_0\rightarrow M_1\rightarrow M_2\rightarrow\ldots$,
then we can build a sequence $N\equiv N_0\rightarrow N_1\rightarrow N_2\rightarrow\ldots$
in a coinductive way:
\begin{varitemize}
  \item
  If $M_1\equiv N$, then we define $N_i$ to be $M_{i+1}$ for every $i\geq 1$.
  \item
  If $M_1\not\equiv N$ then by proposition~\ref{prop:diamprop}
  there is $N_1$ such that $M_1\rightarrow N_1$ and $N_0\rightarrow N_1$.
\end{varitemize}
Now, we prove by induction on $n$ that if $M\rightarrow^n N$, with 
$N$ normal form, then $M$ is strongly normalizing. If $n=0$, then
$M$ is a normal form, then strongly normalizing. If $n\geq 1$, assume,
by way of contraddiction, that $M$ is not strongly normalizing.
Let $L$ be a term such that $M\rightarrow L\rightarrow^{n-1} N$.
By the above observation, $L$ cannot be strongly normalizing, 
but this goes against the induction hypothesis. This concludes
the proof.
\end{proof}}{}
But we can go even further: in this setting, the number of beta-steps
to the normal form is invariant from the evaluation strategy:
\condinc{
\begin{lemma}[Parametrical Diamond Property]\label{lemma:gendiam}
If $M\rightarrow^n N$ and $M\rightarrow^m L$ then
there is a term $P$ such that $N\rightarrow^l P$ and $L\rightarrow^k P$
where $l\leq m$, $k\leq n$ and $n+l=m+k$. 
\end{lemma}
\begin{proof}
We will proceed by induction
on $n+m$. If $n+m=0$, then $P$ will be $N\equiv L\equiv M$.
If $n+m>0$ but either $n=0$ or $m=0$, the thesis easily
follows. So, we can assume both $n>0$ and $m>0$.
Let now $Q$ and $R$ such that
$M\rightarrow Q\rightarrow^{n-1} N$ and
$M\rightarrow R\rightarrow^{m-1} L$. From
proposition~\ref{prop:diamprop}, we can distinguish
two cases:
\begin{varitemize}
  \item
  $Q\equiv R$, By induction hypothesis, we know 
  there is $T$ such that $N\rightarrow^{l} T$
  and $L\rightarrow^{k} T$, where $l\leq m-1\leq m$,
  and $k\leq n-1\leq n$. Moreover $(n-1)+l=(m-1)+k$,
  which yields $n+l=m+k$.
  \item
  There is $T$ with $Q\rightarrow T$ and $R\rightarrow T$. By induction
  hypothesis, there are two terms $U,W$ such that
  $T\rightarrow^i U$, $T\rightarrow^j W$,
  $N\rightarrow^p U$, $L\rightarrow^q W$,
  where $i\leq n-1$, $j\leq m-1$, 
  $p,q\leq 1$, $n-1+p=1+i$ and $m-1+q=1+j$. 
  By induction hypothesis,
  there is $P$ such that $U\rightarrow^r P$ 
  and $W\rightarrow^s P$, where $r\leq j$,
  $s\leq i$ and $r+i=s+j$. But, summing up, this
  implies
  \begin{eqnarray*}
  N&\rightarrow^{p+r}& P\\
  L&\rightarrow^{q+s}& P\\
  p+r&\leq&1+j\leq 1+m-1=m\\
  q+s&\leq&1+i\leq 1+n-1=n\\
  p+r+n&=&(n-1+p)+1+r=1+i+1+r\\
       &=&2+r+i=2+s+j=1+1+j+s\\
       &=&1+m-1+q+s=q+s+n
  \end{eqnarray*}
\end{varitemize} 
This concludes the proof.
\end{proof}}{}
\begin{proposition}~\label{prop:unicitynf}
For every term $M$, there are at most one
normal form $N$ and one integer $n$ such
that $M\rightarrow^n N$.
\end{proposition}
\condinc{
\begin{proof}
Suppose $M\rightarrow^n N$ and $M\rightarrow^m L$, with
$N$ and $L$ normal forms. Then, by lemma~\ref{lemma:gendiam},
there are $P,k,l$ such that
$N\rightarrow^l P$ and $L\rightarrow^k P$ and
$n+l=m+k$. But since $N$ and $L$ are normal forms,
$P\equiv N$, $P\equiv L$ and $l=k=0$, which yields
$N\equiv L$ and $n=m$.
\end{proof}}{}
\section{An Abstract Time Measure}
We can now define an abstract time measure and prove
a diamond property for it. Intuitively, every beta-step
will be endowed with a positive integer cost bounding 
the difference (in size) between the reduct and
the redex.\par
\begin{definition}
\begin{varitemize}
\item
Concatenation of $\alpha,\beta\in\N^*$ is
simply denoted as $\alpha\beta$.
\item
$\timearrownp$ will denote a subset of $\Lambda\times\N^*\times\Lambda$. 
In the following, we will write $M\timearrow{\alpha}N$ standing
for $(M,\alpha,N)\in\timearrownp$. The definition
of $\timearrownp$ (in SOS-style) is the following:
$$
\begin{array}{ccccc}
\infer{M\timearrow{\varepsilon}M}{} &\hspace{1cm}&
\infer{M\timearrow{(n)}N}{M\rightarrow N & n=\max\{1,\length{N}-\length{M}\}} &\hspace{1cm}&
\infer{M\timearrow{\alpha\beta}L}{M\timearrow{\alpha}N & N\timearrow{\beta}L}
\end{array}
$$
Observe we charge $\max\{1,\length{N}-\length{M}\}$ for every step $M\rightarrow N$. In this
way, the cost of a beta-step will always be positive.
\item
Given $\alpha=(n_1,\ldots,n_m)\in\N^*$, define $\truelength{\alpha}=\sum_{i=1}^m n_i$.
\end{varitemize}
\end{definition}
\condinc{
The confluence properties we proved in the previous section can be lifted to this new notion
on weighted reduction.
}{
The result of Proposition~\ref{prop:unicitynf} can be lifted to this new notion
on weighted reduction.
}
\condinc{ 
\begin{proposition}[Diamond Property Revisited]\label{prop:diamproprev}
If $M\timearrow{(n)} N$ and $M\timearrow{(m)} L$, then
either $N\equiv L$ or there is $P$ such that 
$N\timearrow{(m)} P$ and $L\timearrow{(n)} P$.
\end{proposition}
\begin{proof}
We can proceed as in Proposition~\ref{prop:diamprop}.
Observe that if $M\timearrow{\alpha} N$, then
$ML\timearrow{\alpha} NL$ and $LM\timearrow{\alpha} LN$.
We go by induction on the structure of
$M$. Clearly, $M$ cannot be a variable 
nor an abstraction so $M\equiv QR$.
We can distinguish five cases:
\begin{varitemize}
  \item
  If $Q\equiv\lambda x.T$ and $R$ is a
  value, then $N\equiv L\equiv T\{x/R\}$,
  because $R$ is a variable or an
  abstraction.
  \item
  If
  $N\equiv TR$ and
  $L\equiv UR$, where
  $Q\timearrow{(n)} T$ and
  $Q\timearrow{(m)} U$, then we 
  can apply the induction hypothesis, obtaining that
  $T\timearrow{(m)} W$ and
  $U\timearrow{(n)} W$. This, in turn, implies
  $N\timearrow{(m)}WR$ and $L\timearrow{(n)}WR$.
  \item
  Similarly, if $N\equiv QT$ and
  $L\equiv QU$, where $R\timearrow{(n)}T$ and
  $R\timearrow{(m)}U$, then we 
  can apply the induction hypothesis.   
  \item
  If $N\equiv QT$ and
  $L\equiv UR$, where 
  $R\timearrow{(n)} T$ and
  $Q\timearrow{(m)} U$, then
  $N\timearrow{(m)} UT$ and
  $L\timearrow{(n)} UT$.
  \item
  Similarly, if $N\equiv UR$ and
  $L\equiv QT$, where 
  $R\timearrow{(n)} T$ and
  $Q\timearrow{(m)} U$, then
  $N\timearrow{(m)} UT$ and
  $L\timearrow{(n)} UT$.  
\end{varitemize}
This concludes the proof.
\end{proof}
\begin{lemma}[Parametrical Diamond Property Revisited]\label{lemma:gendiamrev}
If $M\timearrow{\alpha}N$ and $M\timearrow{\beta}L$, then
there is a term $P$ such that $N\timearrow{\gamma}P$ and $L\timearrow{\delta} P$
where $\truelength{\alpha\gamma}=\truelength{\beta\delta}$. 
\end{lemma}
\begin{proof}
We proceed by induction
on $\alpha\beta$. If $\alpha=\beta=\varepsilon$, then $P$ will be $N\equiv L\equiv M$.
If $\alpha\beta\neq 0$ but either $\alpha=\varepsilon$ or $\beta=\varepsilon$ , the thesis easily
follows. So, we can assume both $\alpha\neq\varepsilon$ and $\beta\neq\varepsilon$.
Let now $Q$ and $R$ such that
$M\timearrow{(n)} Q\timearrow{\rho} N$ and
$M\timearrow{(m)} R\timearrow{\delta} L$. From
Proposition~\ref{prop:diamproprev}, we can distinguish
two cases:
\begin{varitemize}
  \item
  $Q\equiv R$ (and $m=n$). By induction hypothesis, we know 
  there is $T$ such that $N\timearrow{\gamma} T$
  and $L\timearrow{\delta} T$, where $\truelength{\rho\gamma}=\truelength{\sigma\delta}$,
  which yields $\truelength{\alpha\gamma}=\truelength{\beta\delta}$.
  \item
  There is $T$ with $Q\timearrow{(m)} T$ and $R\timearrow{(n)} T$. By induction
  hypothesis, there are two terms $U,W$ such that
  $T\timearrow{\xi} U$, $T\timearrow{\eta} W$,
  $N\timearrow{\theta} U$, $L\timearrow{\mu} W$,
  where $\truelength{\rho\theta}=\truelength{(m)\xi}$
  and $\truelength{\sigma\mu}=\truelength{(n)\eta}$. 
  By induction hypothesis,
  there is $P$ such that $U\timearrow{\nu} P$ 
  and $W\timearrow{\tau} P$, where $\truelength{\xi\nu}=\truelength{\eta\tau}$. 
  But, summing up, this
  implies
  \begin{eqnarray*}
  N&\timearrow{\theta\nu}& P\\
  L&\timearrow{\eta\tau}& P\\
  \truelength{\alpha\theta\nu}&=&\truelength{(n)\rho\theta\nu}=\truelength{(n)(m)\xi\nu}=\\
     &=&\truelength{(m)(n)\xi\nu}=\truelength{(m)(n)\eta\tau}=\truelength{(m)\sigma\mu\tau}=\\
     &=&\truelength{\beta\mu\tau}
  \end{eqnarray*}
\end{varitemize} 
This concludes the proof.
\end{proof}}{}
\begin{proposition}\label{prop:unicitynftm}
For every term $M$, there are at most one
normal form $N$ and one integer $n$ such
that $M\timearrow{\alpha}N$ and $\truelength{\alpha}=n$.
\end{proposition}
\condinc{
\begin{proof}
Suppose $M\timearrow{\alpha} N$ and $M\timearrow{\beta}L$, with
$N$ and $L$ normal forms. Then, by Lemma~\ref{lemma:gendiam},
there are $P,\gamma,\delta$ such that
$N\timearrow{\gamma} P$ and $L\timearrow{\delta} P$ and
$\truelength{\alpha\gamma}=\truelength{\beta\delta}$. But since 
$N$ and $L$ are normal forms,
$P\equiv N$, $P\equiv L$ and $\gamma=\delta=\varepsilon$, which yields
$N\equiv L$ and $\truelength{\alpha}=\truelength{\beta}$.
\end{proof}}{}
We are now ready to define the abstract time measure which is the core of the paper.
\begin{definition}[Difference cost model]
If 
$M\timearrow{\alpha}N$, where $N$ is a normal form, then
$\timef{M}$ is $\truelength{\alpha}+\length{M}$. If $M$
diverges, then $\timef{M}$ is infinite.
\end{definition}
Observe that this is a good definition, in view of 
Proposition~\ref{prop:unicitynftm}. In other words, showing
$M\timearrow{\alpha}N$ suffices to prove 
$\timef{M}=\truelength{\alpha}+\length{M}$. This will be particularly
useful in the following section.

As an example, consider again the term $\underline{n}\;\underline{2}$ we discussed in the introduction. It reduces to normal form in one step, because we do not reduce under
the abstraction. To force reduction, consider $E\equiv \underline{n}\;\underline{2}\; c$, where $c$ is a free variable; $E$ reduces to 
\[
\lambda y_n.(\lambda y_{n-1}\ldots(\lambda y_2.(\lambda y_1. c^2 y_1)^2 y_2)^2\ldots)y_n
\]
in $\Theta(n)$ beta steps. However, $\timef{E} = \Theta(2^n)$, since at any step
the size of the term is duplicated.
\section{Simulating Turing Machines}
In this and the following section we will show that the difference cost
model satisfies the polynomial invariance thesis. The present section
shows how to encode Turing machines
into the lambda calculus.\par
We denote by $H$ the term $MM$, where
$$
M\equiv\lambda x.\lambda f.f(\lambda z.xxfz).
$$
$H$ is a call-by-value fixed-point operator: for every $N$, there is $\alpha$ such that
\begin{eqnarray*}
HN&\timearrow{\alpha}&N(\lambda z.HNz)\\
\truelength{\alpha}&=&O(\length{N})
\end{eqnarray*}
The lambda term $H$ provides the necessary computational expressive power
to encode the whole class of computable functions.\par
The simplest objects we need to encode in the lambda-calculus are finite
sets. Elements of any finite set $A=\{a_1,\ldots,a_n\}$ can be encoded
as follows:
$$
\cod{a_i}{A}\equiv\lambda x_1.\ldots.\lambda x_n.x_i
$$
Notice that the above encoding induces a total order
on $A$ such that $a_i\leq a_j$ iff $i\leq j$.\par
Other useful objects are finite strings over an arbitrary alphabet,
which will be encoded using a scheme attributed to Scott~\cite{Wadsworth80}.
Let $\Sigma=\{a_1,\ldots,a_n\}$ be a finite alphabet. A string in
$s\in\Sigma^*$ can be represented by a value
$\cod{s}{\Sigma^*}$ as follows, by induction on $s$:
\begin{eqnarray*}
\cod{\varepsilon}{\Sigma^*}&\equiv&\lambda x_1.\ldots.\lambda x_n.\lambda y.y\\
\cod{a_iu}{\Sigma^*}&\equiv&\lambda x_1.\ldots.
  \lambda x_n\lambda y.x_i\cod{u}{\Sigma^*}
\end{eqnarray*}
Observe that representations of symbols in $\Sigma$
and strings in $\Sigma^*$ depend on the cardinality
of $\Sigma$. In other words, if $u\in\Sigma^*$
and $\Sigma\subset\Delta$, $\cod{u}{\Sigma^*}\neq\cod{u}{\Delta^*}$.
Besides data, we want to be able to encode functions between them. In
particular, the way we have defined numerals lets us 
concatenate two strings in linear time in the underlying lambda calculus.
\condinc{
\begin{lemma}
Given a finite alphabet $\Sigma$, there are
terms $\appendchar{\Sigma}$, $\appendstring{\Sigma}$ and
$\appendreverse{\Sigma}$ such that for every $a\in\Sigma$ 
and $u,v\in\Sigma^*$ there are $\alpha,\beta,\gamma$ such that
\begin{eqnarray*}
\appendchar{\Sigma}\cod{a}{\Sigma}\cod{u}{\Sigma^*}&\timearrow{\alpha}&\cod{au}{\Sigma^*}\\
\appendstring{\Sigma}\cod{u}{\Sigma^*}\cod{v}{\Sigma^*}&\timearrow{\beta}&\cod{uv}{\Sigma^*}\\
\appendreverse{\Sigma}\cod{u}{\Sigma^*}\cod{v}{\Sigma^*}&\timearrow{\gamma}&\cod{u^rv}{\Sigma^*}
\end{eqnarray*}
and
\begin{eqnarray*}
\truelength{\alpha}&=&O(1)\\
\truelength{\beta}&=&O(|u|)\\
\truelength{\gamma}&=&O(|u|)
\end{eqnarray*}
\end{lemma}
\begin{proof}
The three terms are defined as follows:
\begin{eqnarray*}
\appendchar{\Sigma}&\equiv&\lambda x.\lambda y.xM_1\ldots M_{|\Sigma|}y\\
\forall i.M_i&\equiv&\lambda y.\lambda x_1.\ldots.\lambda x_{|\Sigma|}.\lambda w.x_iy\\
\appendstring{\Sigma}&\equiv&H(\lambda x.\lambda y.\lambda z.yN_1\ldots N_{|\Sigma|}(\lambda w.w)z)\\
\forall i.N_i&\equiv&\lambda w.\lambda k.(\lambda h.\lambda x_1.\ldots.\lambda x_{|\Sigma|}.\lambda g.x_ih)(xwk)\\
\appendreverse{\Sigma}&\equiv&H(\lambda x.\lambda y.\lambda z.yP_1\ldots P_{|\Sigma|}(\lambda w.w)z)\\
\forall i.P_i&\equiv&\lambda w.\lambda k.xw(\lambda x_1.\ldots.\lambda x_{|\Sigma|}.\lambda h.x_ik)
\end{eqnarray*}
Observe that
\begin{eqnarray*}
\appendchar{\Sigma}\cod{a_i}{\Sigma}\cod{u}{\Sigma^*}&\timearrow{(1,1)}&
  \cod{a_i}{\Sigma}M_1\ldots M_{|\Sigma|}\cod{u}{\Sigma^*}\\
  &\timearrow{\alpha}&M_i\cod{u}{\Sigma^*}\timearrow{1}\cod{a_iu}{\Sigma^*}
\end{eqnarray*}
where $\alpha$ does not depend on $u$. 
Now, let $R_i$ be $N_i\{\lambda z.\appendstring{\Sigma}z/x\}$. Then, we can proceed by induction:
\begin{eqnarray*}
\appendstring{\Sigma}\cod{\varepsilon}{\Sigma^*}\cod{v}{\Sigma^*}&\timearrow{\alpha}&
  (\lambda y.\lambda z.yR_1\ldots R_{|\Sigma|}(\lambda w.w)z)\cod{\varepsilon}{\Sigma^*}\cod{v}{\Sigma^*}\\
  &\timearrow{(1,1)}& \cod{\varepsilon}{\Sigma^*}R_1\ldots R_{|\Sigma|}(\lambda w.w)\cod{v}{\Sigma^*}\\
  &\timearrow{\beta}& (\lambda w.w)\cod{v}{\Sigma^*}\timearrow{(1)} \cod{v}{\Sigma^*}\\
\appendstring{\Sigma}\cod{a_iu}{\Sigma^*}\cod{v}{\Sigma^*}&\timearrow{\alpha}&
  (\lambda y.\lambda z.yR_1\ldots R_{|\Sigma|}(\lambda w.w)z)\cod{a_iu}{\Sigma^*}\cod{v}{\Sigma^*}\\
  &\timearrow{(1,1)}& \cod{a_iu}{\Sigma^*}R_1\ldots R_{|\Sigma|}(\lambda w.w)\cod{v}{\Sigma^*}\\
  &\timearrow{\gamma}& R_i\cod{u}{\Sigma}\cod{v}{\Sigma^*}\\
  &\timearrow{(1,1,1)}& (\lambda h.\lambda x_1.\ldots.\lambda x_{|\Sigma|}.\lambda g.x_ih)
     (\appendstring{\Sigma}\cod{u}{\Sigma}\cod{v}{\Sigma^*})\\
  &\timearrow{\delta}& (\lambda h.\lambda x_1.\ldots.\lambda x_{|\Sigma|}.\lambda g.x_ih)\cod{uv}{\Sigma^*}\\
  &\timearrow{(1)}& \lambda x_1.\ldots.\lambda x_{|\Sigma|}.\lambda g.x_i\cod{uv}{\Sigma^*}
\end{eqnarray*}
where $\alpha,\beta,\gamma$ do no depend on $u$ and $v$. Finally, let $Q_i$ be 
$P_i\{\lambda z.\appendreverse{\Sigma}z/x\}$. Then, we can proceed by induction:
\begin{eqnarray*}
\appendreverse{\Sigma}\cod{\varepsilon}{\Sigma^*}\cod{v}{\Sigma^*}&\timearrow{\alpha}&
  (\lambda y.\lambda z.yQ_1\ldots Q_{|\Sigma|}(\lambda w.w)z)\cod{\varepsilon}{\Sigma^*}\cod{v}{\Sigma^*}\\
  &\timearrow{(1,1)}& \cod{\varepsilon}{\Sigma^*}Q_1\ldots Q_{|\Sigma|}(\lambda w.w)\cod{v}{\Sigma^*}\\
  &\timearrow{\beta}& (\lambda w.w)\cod{v}{\Sigma^*}\timearrow{(1)} \cod{v}{\Sigma^*}\\
\appendreverse{\Sigma}\cod{a_iu}{\Sigma^*}\cod{v}{\Sigma^*}&\timearrow{\alpha}&
  (\lambda y.\lambda z.yQ_1\ldots Q_{|\Sigma|}(\lambda w.w)z)\cod{a_iu}{\Sigma^*}\cod{v}{\Sigma^*}\\
  &\timearrow{(1,1)}& \cod{a_iu}{\Sigma^*}Q_1\ldots Q_{|\Sigma|}(\lambda w.w)\cod{v}{\Sigma^*}\\
  &\timearrow{\gamma}& Q_i\cod{u}{\Sigma}\cod{v}{\Sigma^*}\\
  &\timearrow{(1,1,1)}& \appendreverse{\Sigma}\cod{u}{\Sigma}\cod{a_iv}{\Sigma^*}\\
  &\timearrow{\delta}& \cod{u^ra_iv}{\Sigma^*}\equiv\cod{(a_iu)^rv}{\Sigma^*}
\end{eqnarray*}
where $\alpha,\beta,\gamma$ do not depend on $u,v$. 
\end{proof}}{}
The encoding of a string depends on the underlying alphabet. As a consequence,
we also need to be able to convert representations for strings in one alphabet to
corresponding representations in a bigger alphabet. This can be done efficiently
in the lambda-calculus.
\condinc{
\begin{lemma}
Given two finite alphabets $\Sigma$ and $\Delta$, 
there are terms
$\convertchar{\Sigma}{\Delta}$ and $\convertstring{\Sigma}{\Delta}$
such that for every $a_0,a_1,\ldots,a_n\in\Sigma$ there are $\alpha$ and $\beta$ with
\begin{eqnarray*}
\convertchar{\Sigma}{\Delta}\cod{a_0}{\Sigma}&\timearrow{\alpha}&\cod{u_0}{\Delta^*}\\
\convertstring{\Sigma}{\Delta}\cod{a_1\ldots a_n}{\Sigma^*}&\timearrow{\beta}&\cod{u_1\ldots u_n}{\Delta^*}\\
\forall i.u_i&=&\left\{
                \begin{array}{ll}
                  a_i & \mbox{if $a_i\in\Delta$}\\
                  \varepsilon & \mbox{otherwise}
		\end{array}
                \right.
\end{eqnarray*}
and
\begin{eqnarray*}
\truelength{\alpha}&=&O(1)\\
\truelength{\beta}&=&O(n)\\
\end{eqnarray*}
\end{lemma}
\begin{proof}
The two terms are defined as follows:
\begin{eqnarray*}
\convertchar{\Sigma}{\Delta}&\equiv&\lambda x.xM_1\ldots M_{|\Sigma|}\\
\forall i.M_i&\equiv&\left\{
                          \begin{array}{ll}
                          \cod{a_i}{\Delta^*} & \mbox{if $a_i\in\Delta$}\\
                          \cod{\varepsilon}{\Delta^*} & \mbox{otherwise}
                          \end{array}
	             \right. \\
\convertstring{\Sigma}{\Delta}&\equiv&H(\lambda x.\lambda y.yN_1\ldots N_{|\Sigma|}(\cod{\varepsilon}{\Delta^*}))\\
\forall i.N_i&\equiv&\left\{
                          \begin{array}{ll}
			    \lambda z.(\lambda w.\lambda x_1.\ldots.\lambda x_{|\Delta|}.
			    \lambda h.x_iw)(xz) & \mbox{if $a_i\in\Delta$}\\
			    \lambda z.xz &\mbox{otherwise}
                          \end{array}
                     \right.
\end{eqnarray*}
Observe that
\begin{eqnarray*}
\convertchar{\Sigma}{\Delta}\cod{a_i}{\Sigma}&\timearrow{(1)}&
  \cod{a_i}{\Sigma}M_1\ldots M_{|\Sigma|}\\
  &\timearrow{\alpha}&\left\{
                   \begin{array}{ll}
                      \cod{a_i}{\Delta^*}& \mbox{if $a_i\in\Delta$}\\
                      \cod{\varepsilon}{\Delta^*}& \mbox{otherwise}
                   \end{array}
                   \right.
\end{eqnarray*}
Let $P_i$ be $N_i\{\lambda z.\convertstring{\Sigma}{\Delta}z/x\}$. Then:
\begin{eqnarray*}
\convertstring{\Sigma}{\Delta}\cod{\varepsilon}{\Sigma^*}&\timearrow{\alpha}&
  (\lambda y.yP_1\ldots P_{|\Sigma|}\cod{\varepsilon}{\Delta^*})\cod{\varepsilon}{\Sigma^*}\\
  &\timearrow{(1)}& \cod{\varepsilon}{\Sigma^*}P_1\ldots P_{|\Sigma|}\cod{\varepsilon}{\Delta^*}\\
  &\timearrow{\beta}& \cod{\varepsilon}{\Delta^*}\\
\convertstring{\Sigma}{\Delta}\cod{a_iu}{\Sigma^*}&\timearrow{\gamma}&
  (\lambda y.yP_1\ldots P_{|\Sigma|}\cod{\varepsilon}{\Delta^*})\cod{a_iu}{\Sigma^*}\\
  &\timearrow{(1)}& \cod{a_iu}{\Sigma^*}P_1\ldots P_{|\Sigma|}\cod{\varepsilon}{\Delta^*}\\
  &\timearrow{\delta}& P_i\cod{u}{\Sigma^*}\\
  &\timearrow{(1,1)}&\left\{
                   \begin{array}{ll}
                   (\lambda w.\lambda x_1.\ldots.\lambda x_{|\Delta|}.
			    \lambda h.x_iw)(\convertstring{\Sigma}{\Delta}\cod{u}{\Sigma^*})& \mbox{if $a_i\in\Delta$}\\
                   \convertstring{\Sigma}{\Delta}\cod{u}{\Sigma^*}& \mbox{otherwise}
                   \end{array}
                   \right.
\end{eqnarray*}
where $\alpha,\beta,\gamma,\delta$ do not depend on $u$. 
\end{proof}}{}
A deterministic Turing machine $\mathcal{M}$ is a tuple $(\Sigma,a_{\mathit{blank}},Q,q_{\mathit{initial}},
q_{\mathit{final}},\delta)$ consisting of:
\begin{varitemize}
  \item
  A finite alphabet $\Sigma=\{a_1,\ldots,a_n\}$;
  \item
  A distinguished symbol $a_{\mathit{blank}}\in\Sigma$, called the \emph{blank symbol};
  \item
  A finite set $Q=\{q_1,\ldots,q_m\}$ of \emph{states};
  \item
  A distinguished state $q_{\mathit{initial}}\in Q$, called the \emph{initial
  state};
  \item
  A distinguished state $q_{\mathit{final}}\in Q$, called the \emph{final
  state}; 
  \item
  A partial \emph{transition function} $\delta:Q\times\Sigma\rightharpoonup 
  Q\times\Sigma\times\{\leftarrow,\rightarrow,\downarrow\}$ such that
  $\delta(q_i,a_j)$ is defined iff $q_i\neq q_{\mathit{final}}$.
\end{varitemize}
A configuration for $\mathcal{M}$ is a quadruple in
$\Sigma^*\times\Sigma\times\Sigma^*\times Q$. 
For example, if $\delta(q_i,a_j)=(q_l,a_k,\leftarrow)$, then $\mathcal{M}$ 
evolves from $(ua_p,a_j,v,q_i)$ to $(u,a_p,a_kv,q_l)$ (and from $(\varepsilon,a_j,v,q_i)$
to $(\varepsilon,a_{\mathit{blank}},a_kv,q_l)$). A configuration like
$(u,a_i,v,q_{\mathit{final}})$ is \emph{final} and cannot evolve.
Given a string $u\in\Sigma^*$, the \emph{initial configuration}
for $u$ is $(\varepsilon,a,u,q_{\mathit{initial}})$
if $u=av$ and $(\varepsilon,a_{\mathit{blank}},\varepsilon,q_{\mathit{initial}})$
if $u=\varepsilon$. The string corresponding to the final
configuration $(u,a_i,v,q_{\mathit{final}})$ is $ua_i v$.\par
A Turing machine $(\Sigma,a_{\mathit{blank}},Q,q_{\mathit{initial}},
q_{\mathit{final}},\delta)$ computes the function $f:\Delta^*\rightarrow\Delta^*$ 
(where $\Delta\subseteq\Sigma$) in time $g:\N\rightarrow\N$ 
iff for every $u\in\Delta^*$, the
initial configuration for $u$ evolves to a final configuration
for $f(u)$ in $g(|u|)$ steps.\par
A configuration $(s,a,t,q)$ of a machine
$\M=(\Sigma,a_{\mathit{blank}},Q,q_{\mathit{initial}},
q_{\mathit{final}},\delta)$ is represented by
the term
$$
\cod{(u,a,v,q)}{\M}\equiv\lambda x.x
\cod{u^r}{\Sigma^*}\;\cod{a}{\Sigma}\;\cod{v}{\Sigma^*}\;\cod{q}{Q}
$$
We now encode a Turing machine 
$\M=(\Sigma,a_{\mathit{blank}},Q,q_{\mathit{initial}},
q_{\mathit{final}},\delta)$ in the lambda-calculus.
Suppose $\Sigma=\{a_1,\ldots,a_{|\Sigma|}\}$
and $Q=\{q_1,\ldots,q_{|Q|}\}$
We proceed by building up three lambda terms:
\begin{varitemize} 
  \item
    First of all, we need to be able to build the initial
    configuration for $u$ from $u$ itself. This can be done in linear
    time.
  \item
    Then, we need to extract a string from a final configuration
    for the string. This can be done in linear time, too.
  \item
    Most importantly, we need to be able to simulate the transition function of
    $\M$, i.e. compute a final configuration
    from an initial configuration (if it exists). This
    can be done with cost proportional to the number of
    steps $\M$ takes on the input. 
\end{varitemize}
\condinc{The following three lemmas formalize the above intuitive
argument:
\begin{lemma}
Given a Turing machine $\M=(\Sigma,a_{\mathit{blank}},Q,q_{\mathit{initial}},
q_{\mathit{final}},\delta)$ and an alphabet $\Delta\subseteq\Sigma$ there is a term 
$\init{\M}{\Delta}$ such that for every $u\in\Delta^*$, 
$\init{\M}{\Delta}\cod{u}{\Delta^*}\timearrow{\alpha}\cod{C}{\M}$
where $C$ is the initial configuration for $u$ and $\truelength{\alpha}=O(|u|)$.
\end{lemma}
\begin{proof}
$\init{M}{\Delta}$ is defined as
$$
H(\lambda x.\lambda y.yM_1\ldots M_{|\Delta|}N)
$$
where 
\begin{eqnarray*}
  N&\equiv&\cod{(\varepsilon,a_\mathit{blank},\varepsilon,q_\mathit{initial})}{\M}\\
  M_i&\equiv&\lambda z.(xz)(\lambda u.\lambda a.\lambda v.\lambda q.\lambda w.(\lambda x.x u\cod{a_i}{\Sigma}
      wq)(\appendchar{\Sigma}a v))
\end{eqnarray*}
Let $P_i$ be $M_i\{\lambda z.\init{\M}{\Delta}z/x\}$. Then 
\begin{eqnarray*}
\init{\mathcal{M}}{\Delta}\cod{\varepsilon}{\Delta^*}&\timearrow{\alpha}&(\lambda y.yP_1\ldots P_{|\Delta|}N)
  \cod{\varepsilon}{\Delta^*}\\
&\timearrow{(1)}&\cod{\varepsilon}{\Delta^*}P_1\ldots P_{|\Delta|}N\\
&\timearrow{\beta}&N\equiv\cod{(\varepsilon,a_\mathit{blank},\varepsilon,q_\mathit{initial})}{\M}\\
\init{\mathcal{M}}{\Delta}\cod{a_iu}{\Delta^*}&\timearrow{\alpha}&(\lambda y.yP_1\ldots P_{|\Delta|}N)
  \cod{a_iu}{\Delta^*}\\
&\timearrow{(1)}&\cod{a_iu}{\Delta^*}P_1\ldots P_{|\Delta|}N\\
&\timearrow{\beta}&P_i\cod{u}{\Delta^*}\\
&\timearrow{(1)}&(\init{M}{\Delta}\cod{u}{\Delta^*})
(\lambda u.\lambda a.\lambda v.\lambda q.\lambda w.(\lambda x.x u\cod{a_i}{\Sigma}wq)(\appendchar{\Sigma}a v))\\
&\timearrow{\gamma}&\cod{D}{\M}
(\lambda u.\lambda a.\lambda v.\lambda q.\lambda w.(\lambda x.x u\cod{a_i}{\Sigma}wq)(\appendchar{\Sigma}a v))
\end{eqnarray*}
where $\alpha,\beta$ do not depend on $u$ and $D$ is and initial configuration for $u$. Clearly
$$
\cod{D}{\M}
(\lambda u.\lambda a.\lambda v.\lambda q.\lambda w.(\lambda x.x u\cod{a_i}{\Sigma}wq)(\appendchar{\Sigma}a v))
\timearrow{(1,1,1,1,1)}\cod{E}{\M}
$$
where $E$ is an initial configuration for $a_iu$. 
\end{proof}
\begin{lemma}
Given a Turing machine $\M=(\Sigma,a_{\mathit{blank}},Q,q_{\mathit{initial}},
q_{\mathit{final}},\delta)$ and for every alphabet $\Delta$, there is
a term $\final{\M}{\Delta}$ such that for every final
configuration $C$ for $u_1\ldots u_n$ there is $\alpha$ such that
$\final{\M}{\Delta}\cod{C}{\M}\timearrow{\alpha}\cod{v_1\ldots v_n}{\Delta^*}$, 
$\truelength{\alpha}=O(n)$ and
$$
\forall i.v_i=\left\{
                \begin{array}{ll}
                  u_i & \mbox{if $u_i\in\Delta$}\\
                  \varepsilon & \mbox{otherwise}
		\end{array}
                \right.
$$
\end{lemma}
\begin{proof}
$\final{\M}{\Delta}$ is defined as
$$
\lambda x.x(\lambda u.\lambda a.\lambda v.\lambda q.
   \appendreverse{\Sigma}
   (\convertstring{\Sigma}{\Delta}u)
   (\appendstring{\Sigma}
      (\convertchar{\Sigma}{\Delta}a)
      (\convertstring{\Sigma}{\Delta}v))
$$ 
Consider an arbitrary final configuration $\cod{(u,a,v,q_\mathit{final})}{\M}$.
Then:
\begin{eqnarray*}
  & &\final{\M}{\Delta}\cod{(u,a,v,q_\mathit{final})}{\M}\\
  &\timearrow{(1,1,1,1,1)}&
   \appendreverse{\Sigma}
   (\convertstring{\Sigma}{\Delta}\cod{u}{\Sigma^*})
   (\appendstring{\Sigma}
      (\convertchar{\Sigma}{\Delta}\cod{a}{\Sigma})
      (\convertstring{\Sigma}{\Delta}{\cod{v}{\Sigma^*}}))\\
  &\timearrow{\alpha}&
  \appendreverse{\Sigma}
   (\cod{u}{\Delta^*})(\appendstring{\Delta}
      (\cod{a}{\Delta^*})
      (\cod{v}{\Delta^*}))\\
  &\timearrow{\beta}&
  \appendreverse{\Sigma}
   \cod{u}{\Delta^*}\cod{av}{\Delta^*}\\
 &\timearrow{\gamma}&
   \cod{u^rav}{\Delta^*}
\end{eqnarray*}
where $\alpha=O(|u|+|v|)$, $\beta$ does not depend on $u,v$ and
$\gamma=O(|u|)$.
\end{proof}
\begin{lemma}
Given a Turing machine $\M=(\Sigma,a_{\mathit{blank}},Q,q_{\mathit{initial}},
q_{\mathit{final}},\delta)$, there is a term $\trans{\M}$ such that 
for every configuration $C$: 
\begin{varitemize}
  \item
  If $D$ is a final configuration reachable from $C$ in $n$ steps,
  then there exists $\alpha$ such that $\trans{\M}\cod{C}{\M}\timearrow{\alpha}\cod{D}{\M}$;
  moreover $\truelength{\alpha}=O(n)$;
  \item
  The term $\trans{\M}\cod{C}{\M}$ diverges if there is
  no final configuration reachable from $C$.
\end{varitemize}
\end{lemma}
\begin{proof}
$\trans{\M}$ is defined as
$$
H(\lambda x.\lambda y.y(\lambda u.\lambda a.\lambda v.\lambda q.q(M_1\ldots M_{|Q|})uav))
$$
where 
\begin{eqnarray*}
  \forall i.M_i&\equiv&\lambda u.\lambda a.\lambda v.a(N_i^1\ldots N_i^{|\Sigma|})uv\\
  \forall i,j.N_i^j&\equiv&
   \left\{
   \begin{array}{ll}
   \lambda u.\lambda v.\lambda x.xu\cod{a_j}{\Sigma}v\cod{q_i}{Q}
      &\mbox{if $q_i=q_\mathit{final}$}\\
   \lambda u.\lambda v.x(\lambda z.zu\cod{a_k}{\Sigma}v\cod{q_l}{Q}) 
      &\mbox{if $\delta(q_i,a_j)=(q_l,a_k,\downarrow)$}\\
   \lambda u.\lambda v.x(uP_1\ldots P_{|\Sigma|}P(\appendchar{\Sigma}
                                               \cod{a_k}{\Sigma}v)
                                               \cod{q_l}{Q})
      &\mbox{if $\delta(q_i,a_j)=(q_l,a_k,\leftarrow)$}\\ 
   \lambda u.\lambda v.x(vR_1\ldots R_{|\Sigma|}R(\appendchar{\Sigma}
                                               \cod{a_k}{\Sigma}u)
                                               \cod{q_l}{Q})
      &\mbox{if $\delta(q_i,a_j)=(q_l,a_k,\rightarrow)$}
   \end{array}
    \right.\\
  \forall i.P_i&\equiv&\lambda u.\lambda v.\lambda q.\lambda x.xu\cod{a_i}{\Sigma}vq\\
   P&\equiv&\lambda v.\lambda q.\lambda x.x\cod{\varepsilon}{\Sigma^*}
      \cod{a_\mathit{blank}}{\Sigma}vq\\
  \forall i.R_i&\equiv&\lambda v.\lambda u.\lambda q.\lambda x.xu\cod{a_i}{\Sigma}vq\\
   R&\equiv&\lambda u.\lambda q.\lambda x.xu
      \cod{a_\mathit{blank}}{\Sigma}\cod{\varepsilon}{\Sigma^*}q
\end{eqnarray*}
To prove the thesis, it suffices to show that
$$
\trans{\M}\cod{C}{\M}\timearrow{\beta}\trans{\M}\cod{E}{\M}
$$
where $E$ is the next configuration reachable from $C$
and $\beta$ is bounded by a constant independent of $C$. We need a number of
abbreviations:
$$
\begin{array}{lll}
\forall i.Q_i\equiv M_i\{\lambda z.\trans{\M}z/x\} & 
\forall i.U_i\equiv P_i\{\lambda z.\trans{\M}z/x\} &
\forall i.W_i\equiv R_i\{\lambda z.\trans{\M}z/x\}\\
\forall i,j.T_i^j\equiv N_i^j\{\lambda z.\trans{\M}z/x\} & 
U\equiv P\{\lambda z.\trans{\M}z/x\} &
W\equiv R\{\lambda z.\trans{\M}z/x\}
\end{array}
$$
Suppose $C=(u,a_j,v,q_i)$. Then
\begin{eqnarray*}
\trans{\M}\cod{C}{\M}&\timearrow{\gamma}&\cod{q_i}{Q}Q_1\ldots Q_{|Q|}
  \cod{u}{\Sigma^*}\cod{a_j}{\Sigma}\cod{v}{\Sigma^*}\\
&\timearrow{\delta}&Q_i\cod{u}{\Sigma^*}\cod{a_j}{\Sigma}\cod{v}{\Sigma^*}\\
&\timearrow{(1,1,1)}&\cod{a_j}{\Sigma}T_i^1\ldots T_i^j\cod{u}{\Sigma^*}\cod{v}{\Sigma^*}\\
&\timearrow{\rho}&T_i^j\cod{u}{\Sigma^*}\cod{v}{\Sigma^*}
\end{eqnarray*}
where $\gamma,\delta,\rho$ do not depend on $C$. Now, consider the following 
four cases, depending on the value of $\delta(q_i,a_j)$:
\begin{varitemize}
   \item
   If $\delta(q_i,a_j)$ is undefined, then $q_i=q_\mathit{final}$ and, by
   definition $T_i^j\equiv\lambda u.\lambda v.\lambda x.xu\cod{a_j}{\Sigma}v\cod{q_i}{Q}$.
   As a consequence,
  \begin{eqnarray*}
   T_i^j\cod{u}{\Sigma^*}\cod{v}{\Sigma^*}&\timearrow{(1,1)}&
  \lambda x.x\cod{u}{\Sigma^*}\cod{a_j}{\Sigma}\cod{v}{\Sigma^*}\cod{q_i}{Q}\\
   &\equiv&\cod{(u,a_j,v,q_i)}{\M}
  \end{eqnarray*} 
   \item
   If $\delta(q_i,a_j)=(q_l,a_k,\downarrow)$, then
   $
   T_i^j\equiv\lambda u.\lambda v.(\lambda z.\trans{\M}z)(\lambda z.zu\cod{a_k}{\Sigma}v\cod{q_l}{Q}).
   $
   As a consequence,
   \begin{eqnarray*}
   T_i^j\cod{u}{\Sigma^*}\cod{v}{\Sigma^*}&\timearrow{(1,1)}&
   (\lambda z.\trans{\M}z)(\lambda z.z\cod{u}{\Sigma^*}\cod{a_k}{\Sigma}\cod{v}{\Sigma^*}\cod{q_l}{Q})\\
   &\timearrow{(1)}&\trans{\M}(\lambda z.z\cod{u}{\Sigma^*}\cod{a_k}{\Sigma}\cod{v}{\Sigma^*}\cod{q_l}{Q})\\
   &\equiv&\trans{\M}\cod{E}{\M}
   \end{eqnarray*} 
   \item
   If $\delta(q_i,a_j)=(q_l,a_k,\leftarrow)$, then
   $$
   \lambda u.\lambda v.x(uU_1\ldots U_{|\Sigma|}U(\appendchar{\Sigma}
                                               \cod{a_j}{\Sigma}v)
                                               \cod{q_l}{Q}).
   $$
   As a consequence,
   \begin{eqnarray*}
   T_i^j\cod{u}{\Sigma^*}\cod{v}{\Sigma^*}&\timearrow{(1,1)}&
     (\lambda z.\trans{\M}z)(\cod{u}{\Sigma^*}U_1\ldots U_{|\Sigma|}U(\appendchar{\Sigma}
     \cod{a_j}{\Sigma}\cod{v}{\Sigma^*})\cod{q_l}{Q})
   \end{eqnarray*} 
Now, if $u$ is $\varepsilon$, then
   \begin{eqnarray*}
     &&(\lambda z.\trans{\M}z)(\cod{u}{\Sigma^*}U_1\ldots U_{|\Sigma|}U(\appendchar{\Sigma}
     \cod{a_j}{\Sigma}\cod{v}{\Sigma^*})\cod{q_l}{Q})\\
     &\timearrow{\eta}&
      (\lambda z.\trans{\M}z)U(\appendchar{\Sigma}\cod{a_j}{\Sigma}\cod{v}{\Sigma^*})\cod{q_l}{Q}\\
     &\timearrow{\xi}&(\lambda z.\trans{\M}z)U(\cod{a_jv}{\Sigma^*})\cod{q_l}{Q})\\
     &\timearrow{(1,1)}&(\lambda z.\trans{\M}z)\cod{(\varepsilon,a_\mathit{blank},a_kv,q_l)}{\M}\\
     &\timearrow{(1)}&\trans{\M}\cod{(\varepsilon,a_\mathit{blank},a_kv,q_l)}{\M}
   \end{eqnarray*} 
where $\eta,\xi$ do not depend on $C$. If $u$ is $ta_p$, then
   \begin{eqnarray*}
     &&(\lambda z.\trans{\M}z)(\cod{u^r}{\Sigma^*}U_1\ldots U_{|\Sigma|}U(\appendchar{\Sigma}
     \cod{a_j}{\Sigma}\cod{v}{\Sigma^*})\cod{q_l}{Q})\\
     &\timearrow{\pi}&
      (\lambda z.\trans{\M}z)U_p\cod{t^r}{\Sigma^*}
      (\appendchar{\Sigma}\cod{a_j}{\Sigma}\cod{v}{\Sigma^*})\cod{q_l}{Q}\\
     &\timearrow{\theta}&(\lambda z.\trans{\M}z)U_p\cod{t^r}{\Sigma^*}(\cod{a_kv}{\Sigma^*})\cod{q_l}{Q})\\
     &\timearrow{(1,1)}&(\lambda z.\trans{\M}z)\cod{(t,a_j,a_kv,q_l)}{\M}\\
     &\timearrow{(1,1)}&\trans{\M}\cod{(t,a_j,a_kv,q_l)}{\M}
   \end{eqnarray*} 
   where $\pi,\theta$ do not depend on $C$.
   \item
   The case $\delta(q_i,a_j)=(q_l,a_k,\rightarrow)$ can be treated similarly.
\end{varitemize}
This concludes the proof.
\end{proof}}{}
At this point, we can give the main simulation result:
\begin{theorem}
If $f:\Delta^*\rightarrow\Delta^*$ is computed by a Turing machine
$\M$ in time $g$, then there is a term $\function{\M}{\Delta}$
such that for every $u\in\Delta^*$ there is $\alpha$ with
$\function{\M}{\Delta}\cod{u}{\Delta^*}\timearrow{\alpha}\cod{f(u)}{\Delta^*}$
and $\truelength{\alpha}=O(g(|u|))$
\end{theorem}
\condinc{
\begin{proof}
Simply define $\function{\M}{\Delta}\equiv \lambda x.\final{\M}{\Delta}(\trans{\M}(\init{\M}{\Delta}x))$.
\end{proof}}{}
Noticeably, the just described simulation induces a linear overhead: every step of $\M$ corresponds
to a constant cost in the simulation, the constant cost not depending on the input but only on
$\M$ itself.
\section{Evaluating with Turing Machines}
We informally describe a Turing machine $\mathcal{R}$ computing 
the normal form of 
a given input term, if it exists,
and diverging otherwise. If $M$ is the input term, $\mathcal{R}$ takes time
$O((\timef{M})^4)$.\par
First of all, let us observe that the usual notation for terms does not 
take into account the complexity of handling variables, and
substitutions. We introduce a notation in the style 
of deBruijn~\cite{deBruijn72}, with binary strings representing 
occurrences of variables. In this way, terms can be denoted by
finite strings in a finite alphabet.
\begin{definition}
\begin{varitemize}
  \item
  The alphabet $\Theta$ is $\{\lambda,@,0,1,\btr \}$.
  \item
  To each lambda term $M$ we can associate a string $M^\#\in\Theta^+$
  in the standard deBruijn way, writing $@$ for (prefix) application. 
  For example, if $M\equiv (\lambda x.xy)(\lambda x.\lambda y.\lambda z.x)$,
  then $M^\#$ is
  $$
  @\lambda @\btr 0\btr\lambda\lambda\lambda\btr 1 0
  $$
  In other words, free occurrences of variables are translated into $\btr$, while
  bounded occurrences of variables are translated into $\btr s$, where $s$ is the
  binary representation of the deBruijn index for that occurrence.
  \item
  The \emph{true length} $\truelength{M}$ of a term $M$ is the length of $M^\#$.
\end{varitemize}
\end{definition}
Observe that $\truelength{M}$ grows more than linearly on $\length{M}$:
\begin{lemma}
For every term $M$, $\truelength{M}=O(\length{M}\log\length{M})$.
There is a sequence $\{M_n\}_{n\in\N}$ such that
$\length{M_n}=\Theta(n)$, while $\truelength{M_n}=
\Theta(\length{M_n}\log\length{M_n})$.
\end{lemma}
\condinc{
\begin{proof}
Consider the following statement: for every $M$, the string
$M^\#$ contains at most $2|M|-1$ characters from $\{\lambda,@\}$
and at most $|M|$ blocks of characters from $\{0,1,\btr \}$,
the length of each of them being at most $1+\lceil\log_2|M|\rceil$.
We proceed by induction on $M$:
\begin{varitemize}
  \item
    If $M$ is a variable $x$, then $M^\#$ is $\btr$. The thesis
    is satisfied, because $\length{M}=1$.
  \item
    If $M$ is $\lambda x.N$, then $M^\#$ is $\lambda u$, where
    $u$ is obtained from $N^\#$ by replacing some blocks
    in the form $\btr$ with $\btr s$, where $\length{s}$ is
    at most $\lceil\log_2|M|\rceil$. As a consequence, the thesis
    remains satisfied.
  \item
    If $M$ is $NL$, then $M^\#$ is $@ N^\# L^\#$ and the thesis
    remains satisfied.
\end{varitemize}
This proves $\truelength{M}=O(\length{M}\log\length{M})$. For the second part, define
$$
M_n\equiv \lambda x.\overbrace{\lambda y.\ldots.\lambda y}^{\mbox{$n$ times}}.
\overbrace{x\ldots x}^{\mbox{$n+1$ times}}
$$
Clearly,
$$
M_n^\#\equiv\overbrace{\lambda\ldots\lambda}^{\mbox{$n+1$ times}}
\overbrace{@\btr u\ldots @\btr u}^{\mbox{$n$ times}}\btr u 
$$
where $u$ is the binary coding of $n$ (so $\length{u}=\Theta(\log n)$).
As a consequence:
\begin{eqnarray*}
\length{M}&=&3n+3=\Theta(n);\\
\truelength{M}&=&\length{M^\#}=3n+3+n\length{u}=\Theta(n\log n).
\end{eqnarray*}
This concludes the proof.
\end{proof}}{}
$\mathcal{R}$ has nine tapes, expects its input to be 
in the first tape and writes the output on the same tape. The tapes will be referred
to as $\Current$ (the first one), $\Preredex$, $\Function$, $\Argument$, 
$\Postredex$, $\Reduct$, $\StackTerm$, $\StackRedex$, $\Counter$. 
$\mathcal{R}$ operates by iteratively performing the following
four steps:
\begin{numlist}
  \item\label{step:first}
  First of all, $\mathcal{R}$ looks for redexes in the term stored
  in $\Current$ (call it $M$), by scanning it. The functional part 
  of the redex will be put
  in $\Function$ while its argument is copied into $\Argument$. Everything
  appearing before (respectively, after) the redex is copied into
  $\Preredex$ (respectively, in $\Postredex$). If there is no redex in $M$, 
  then $\mathcal{R}$ halts. For example, consider the term
  $(\lambda x.\lambda y.xyy)(\lambda z.z)(\lambda w.w)$
  which becomes $@@\lambda\lambda @@\btr 1\btr 0\btr 0\lambda\btr 0\lambda\btr 0$
  in deBruijn notation. Table~\ref{table:afterfirst} summarizes the status 
  of some tapes after this initial step.
  \begin{table}
  \caption{The status of some tapes after step~\ref{step:first}}
  \label{table:afterfirst}
  \begin{center}
  \begin{tabular}{cc}\hline
  $\Preredex$  & $@@$\\
  $\Function$  & $\lambda\lambda @@\btr 1\btr 0\btr 0$\\
  $\Argument$  & $\lambda\btr 0$\\
  $\Postredex$ & $\lambda\btr 0$\\\hline
  \end{tabular}
  \end{center}
  \end{table}
  \item\label{step:second} 
  Then, $\mathcal{R}$ copies the content of $\Function$
  into $\Reduct$, erasing the first occurrence 
  of $\lambda$ and replacing every occurrence of the 
  bounded variable by the content of $\Argument$. 
  In the example, $\Reduct$ becomes $\lambda @@\lambda\btr 0\btr 0\btr 0$.
  \item
  $\mathcal{R}$ replaces the content of $\Current$ with the concatenation
  of $\Preredex$, $\Reduct$ and $\Postredex$ in this particular order.
  In the example, $\Current$ becomes 
  $@\lambda@@\lambda\btr 0\btr 0\btr 0\lambda\btr 0$, 
  which correctly correspond to $(\lambda y.(\lambda z.z)yy)(\lambda w.w)$.
  \item\label{step:last}
  Finally, the content of every tape except $\Current$ is erased.
\end{numlist}
Every time the sequence of steps from~\ref{step:first} to~\ref{step:last} is
performed, the term $M$ in $\Current$ is replaced by another term which is obtained
from $M$ by performing a normalization step. So, $\mathcal{R}$ halts on $M$
if and only if $M$ is normalizing and the output will be the normal form
of $M$.\par
Tapes $\StackTerm$ and $\StackRedex$ are managed in the same way. 
They help keeping track of the structure of a term as it is scanned. 
The two tapes can only contain symbols $A_\lambda$, $F_@$ and $S_@$. 
In particular:
\begin{varitemize}
   \item
   The symbol $A_\lambda$ stands for the argument of an abstraction;
   \item
   the symbol $F_@$ stands for the first argument of an application;
   \item
   the symbol $S_@$ stands for the second argument of an application;
\end{varitemize}
$\StackTerm$ and $\StackRedex$ can only be modified by the usual
stack operations, i.e. by pushing and popping symbols from the top
of the stack. Anytime a new symbol is scanned, the underlying
stack can possibly be modified:
\begin{varitemize}
  \item
  If $@$ is read, then $F_@$ must be pushed on the top of the stack. 
  \item
  If $\lambda$ is read, then $A_\lambda$ must be pushed on the top
  of the stack.
  \item
  If $\btr$ is read, then symbols $S_@$ and $A_\lambda$ must
  be popped from the stack, until we find an occurrence
  of $F_@$ (which must be popped and replaced by $S_@$) or
  the stack is empty.
\end{varitemize}
For example, when scanning the term $@\lambda \btr 0\lambda\btr 0$,
the underlying stack evolves as in table~\ref{table:examplestack}
(the symbol currently being read is underlined).
  \begin{table}
  \caption{How stack evolves while processing $@\lambda \btr 0\lambda \btr 0$}
  \label{table:examplestack}
  \begin{center}
  \begin{tabular}{cc}\hline
  $\underline{@}\lambda \btr \lambda\btr 0$  & $F_@$ \\
  $@\underline{\lambda} \btr 0\lambda\btr 0$  & $F_@ A_\lambda$\\
  $@\lambda \underline{\btr} 0\lambda\btr 0$  & $S_@$\\
  $@\lambda \btr \underline{0}\lambda\btr 0$  & $S_@$\\
  $@\lambda \btr 0\underline{\lambda}\btr 0$  & $S_@ A_\lambda$\\
  $@\lambda \btr 0\lambda\underline{\btr} 0$  & $\varepsilon$\\
  $@\lambda \btr 0\lambda \btr\underline{0}$  & $\varepsilon$\\\hline
  \end{tabular}
  \end{center}
  \end{table}
Now, consider an arbitrary iteration step, where $M$ is reduced
to $N$. We claim that the steps~\ref{step:first} to~\ref{step:last}
can all be performed in $O((\truelength{M}+\truelength{N})^2)$.
\condinc{
The following is an informal argument.
\begin{varitemize}
  \item
  Step~\ref{step:first} can be performed with the help of auxiliary
  tapes $\StackTerm$ and $\StackRedex$. $\Current$ is scanned
  with the help of $\StackTerm$. As soon as $\mathcal{R}$ encounter a 
  $\lambda$ symbol in $\Current$, it treats the subterm  
  in a different way, copying it into $\Function$ with the help of 
  $\StackRedex$.
  When the subterm has been completely processed (i.e. when
  $\StackRedex$ is becomes empty), the machine can
  verify whether or not it is the functional part of a redex. It suffices
  to check the topmost symbol of $\StackTerm$ and the next symbol
  in $\Current$. We are in presence of a redex only if
  the topmost symbol of $\StackTerm$ is $F_@$ and the next
  symbol in $\Current$ is either $\lambda$ or $\btr$.
  Then, $\mathcal{R}$ proceeds as follows: 
  \begin{varitemize}
    \item
    If we are in presence of a redex, then the subterm corresponding
    to the argument is copied into $\Argument$, with the help
    of $\StackRedex$;
    \item
    Otherwise, the content of $\Function$ is moved to
    $\Preredex$ and $\Function$ is completely erased.
  \end{varitemize}
  \item
  Step~\ref{step:second} can be performed with the help
  of $\StackRedex$ and $\Counter$. Initially, $\mathcal{R}$
  simply writes $0$ into $\Counter$, which keeps track of
  $\lambda$-nesting depth of the current symbol 
  (in binary notation) while scanning
  $\Function$. $\StackRedex$ is used in the usual
  way. Whenever we push $A_\lambda$ into $\StackRedex$,
  $\Counter$ is incremented by $1$, while it is decremented
  by $1$ whenever $A_\lambda$ is popped from $\StackRedex$. 
  While scanning $\Function$, $\mathcal{R}$ copies everything 
  into $\Reduct$. If $\mathcal{R}$ encounters a $\btr$, it compares the 
  binary string following it with the actual content
  of $\Counter$. Then it proceeds as follows:
  \begin{varitemize}
    \item
    If they are equal, $\mathcal{R}$ copies to $\Reduct$ the
    entire content of $\Argument$.
    \item
    Otherwise, $\mathcal{R}$ copies to $\Reduct$ the
    representation of the variable occurrences, without
    altering it.
  \end{varitemize}
\end{varitemize}}{}
\begin{lemma}
If $M\rightarrow^n N$, then $n\leq\timef{M}$ and 
$\length{N}\leq\timef{M}$.
\end{lemma}
\begin{proof}
Clear from the definition of $\timef{M}$.
\end{proof}
\begin{theorem}
$\mathcal{R}$ computes the normal form of the
term $M$ in $O((\timef{M})^4)$ steps.
\end{theorem}

\section{Closed Values as a Partial Combinatory Algebra}
If $U$ and $V$ are closed values and $UV$ has
a normal form $W$ (which must be a closed value), 
then we will denote $W$ by
$\app{U}{V}$. In this way, we can give $\Xi$
the status of a partial applicative structure,
which turns out to be a partial combinatory algebra.
The abstract time measure induces a finer
structure on $\Xi$, which we are going to \condinc{illustrate}{sketch} in
this section. In particular, we will be able to show the
existence of certain elements of $\Xi$ having both usual
combinatorial properties as well as bounded behaviour. 
These properties are exploited in~\cite{dallago05fsttcs}, where elements
of $\Xi$ serves as (bounded) realizers in a semantic
framework.\par
In the following, $\timeapp{U}{V}$ is 
simply $\timef{UV}$ (if it exists). Moreover,
$\langle V,U\rangle$ will denote the term $\lambda x.xVU$.\par
First of all, we observe the identity and basic operations on couples take
constant time. For example, there is a term $M_\mathit{swap}$
such that $\app{M_\mathit{swap}}{\langle V,U\rangle}=\langle U,V\rangle$
and $\timeapp{M_\mathit{swap}}{\langle V,U\rangle}=5$.
\condinc{
Formally:
\begin{proposition}[Basic Operators]
There are terms $M_\mathit{id},M_\mathit{swap},M_\mathit{assl},M_\mathit{tens}\in\Xi$ 
and constants $c_\mathit{id}$, $c_\mathit{swap}$, $c_\mathit{assl}$, 
$c^1_\mathit{tens}$ and $c^2_\mathit{tens}$ such that, for every 
$V,U,W\in\Xi$, there is $Y\in\Xi$ such that
\begin{eqnarray*}
\app{M_\mathit{id}}{V}&=&V\\
\app{M_\mathit{swap}}{\langle V,U\rangle}&=&\langle U,V\rangle\\
\app{M_\mathit{assl}}{\langle V,\langle U,W\rangle\rangle}&=&\langle\langle V,U\rangle,W\rangle\\
\app{M_\mathit{tens}}{V}&=&Y\\
\app{Y}{\langle U,W\rangle}&=&\langle \app{V}{U},W\rangle\\
\timeapp{M_\mathit{id}}{V}&\leq&c_\mathit{id}\\
\timeapp{M_\mathit{swap}}{\langle V,U\rangle}&\leq&c_\mathit{swap}\\
\timeapp{M_\mathit{assl}}{\langle V,\langle U,W\rangle\rangle}&\leq& c_\mathit{assl}\\
\timeapp{M_\mathit{tens}}{V}&\leq& c_\mathit{tens}^1\\
\timeapp{Y}{\langle U,W\rangle}&\leq& c_\mathit{tens}^2+\timeapp{V}{U}
\end{eqnarray*}
\end{proposition}
\begin{proof}
First of all, let us define terms:
\begin{eqnarray*}
  M_\mathit{id}&\equiv&\lambda x.x\\
  M_\mathit{swap}&\equiv&\lambda x.x(\lambda y.\lambda w.\lambda z.zwy)\\
  M_\mathit{assl}&\equiv&\lambda x.x(\lambda y.\lambda w.w(\lambda z.\lambda q.\lambda r.r(\lambda s.syz)q))\\
  M_\mathit{tens}&\equiv&\lambda s.\lambda x.x(\lambda y.\lambda w.(\lambda x.\lambda z.zxw)(sy))
\end{eqnarray*}
Now, let us observe that
\begin{eqnarray*}
 M_\mathit{id}V &\timearrow{(1)} V& \\
 M_\mathit{swap}\langle V,U\rangle &\timearrow{(1)}& \langle V,U\rangle (\lambda y.\lambda w.\lambda z.zwy)\\
  &\timearrow{(1)}& (\lambda y.\lambda w.\lambda z.zwy)VU\\
  &\timearrow{(1)}& (\lambda w.\lambda z.zwV)U\\
  &\timearrow{(1)}& (\lambda w.\lambda z.zwV)U\\
  &\timearrow{(1)}& \langle U,V\rangle \\
M_\mathit{assl}\langle V,\langle U,W\rangle\rangle &\timearrow{(1)}&\langle V,\langle U,W\rangle\rangle
  (\lambda y.\lambda w.w(\lambda z.\lambda q.\lambda r.r(\lambda s.syz)q))\\
  &\timearrow{(1)}&(\lambda y.\lambda w.w(\lambda z.\lambda q.\lambda r.r(\lambda s.syz)q))V\langle U,W\rangle\\
  &\timearrow{(1)}&(\lambda w.w(\lambda z.\lambda q.\lambda r.r(\lambda s.sVz)q))\langle U,W\rangle\\
  &\timearrow{(1)}&\langle U,W\rangle(\lambda z.\lambda q.\lambda r.r(\lambda s.sVz)q)\\
  &\timearrow{(1)}&(\lambda z.\lambda q.\lambda r.r(\lambda s.sVz)q)UW\\
  &\timearrow{(1)}&\lambda r.r(\lambda s.sVU)W)\equiv\langle\langle V,U\rangle,W\rangle\\
M_\mathit{tens}V&\timearrow{(1)}&\lambda x.x(\lambda y.\lambda w.(\lambda x.\lambda z.zxw)(Vy))\equiv Y\\
  Y\langle U,W\rangle&\timearrow{(1)}&\langle U,W\rangle(\lambda y.\lambda w.(\lambda x.\lambda z.zxw)(Vy))\\
  &\timearrow{(1)}&(\lambda y.\lambda w.(\lambda x.\lambda z.zxw)(Vy))UW\\
  &\timearrow{(1)}&(\lambda w.(\lambda x.\lambda z.zxw)(VU))W\\
  &\timearrow{(1)}&(\lambda x.\lambda z.zxW)(VU)
\end{eqnarray*}
\end{proof}}{}
There is a term in $\Xi$ which takes as input a pair of terms $\langle V,U\rangle$
and computes the composition of the functions computed by $V$ and $U$. The
overhead is constant, i.e. do not depend on the intermediate result.
\condinc{
\begin{proposition}[Composition]
There are a term $M_\mathit{conc}\in\Xi$ and two constants 
$c^1_\mathit{conc},c^2_\mathit{conc}$ such that, for every 
$V,U,W,Z\in\Xi$, there is $X\in\Xi$ such that:
\begin{eqnarray*}
\app{M_\mathit{conc}}{\langle V,U\rangle}&=&X\\
\app{X}{W}&=&\app{V}{\app{U}{W}}\\
\timeapp{M_\mathit{conc}}{\langle V,U\rangle}&\leq&c_\mathit{conc}^1\\
\timeapp{X}{W}&\leq&c_\mathit{conc}^2+\timeapp{U}{W}+\timeapp{V}{\app{U}{W}}
\end{eqnarray*}
\end{proposition}
\begin{proof}
First of all, let us define term:
$$
M_\mathit{conc}\equiv\lambda x.x(\lambda x.\lambda y.\lambda z.x(yz))
$$
Now, let us observe that
\begin{eqnarray*}
  M_\mathit{conc}\langle V,U\rangle &\timearrow{(1)}&\langle V,U\rangle(\lambda x.\lambda y.\lambda z.x(yz)) \\
  &\timearrow{(1)}&(\lambda x.\lambda y.\lambda z.x(yz))VU\\
  &\timearrow{(1)}&(\lambda y.\lambda z.V(yz))U\timearrow{(1)}\lambda z.V(Uz)\equiv X\\
  XW&\timearrow{(1)}&V(UW)
\end{eqnarray*}
\end{proof}}{}
\condinc{
We need to represent functions which go beyond the realm of linear logic. In particular,
terms can be duplicated, but linear time is needed to do it.
\begin{proposition}[Contraction]
There are a term $M_\mathit{cont}\in\Xi$ and a constant 
$c_\mathit{cont}$ such that, for every $V\in\Xi$:
\begin{eqnarray*}
\app{M_\mathit{cont}}V&=&\langle V,V\rangle\\
\timeapp{M_\mathit{cont}}{V}&\leq&c_\mathit{cont}+\length{V}.
\end{eqnarray*}
\end{proposition}
\begin{proof}
First of all, let us define term:
$$
  M_\mathit{cont}\equiv\lambda x.\lambda y.yxx
$$
Now, let us observe that
$$
  M_\mathit{cont}V\timearrow{(n)}\langle V,V\rangle,
$$
where $n\leq\truelength{V}$. 
\end{proof}}{
We need to represent functions which go beyond the realm of linear logic. In particular,
terms can be duplicated, but linear time is needed to do it: there is
a term $M_\mathit{cont}$ such that $\app{M_\mathit{cont}}V=\langle V,V\rangle$
and $\timeapp{M_\mathit{cont}}{V}=O(\length{V})$.}
From a complexity viewpoint, what is most interesting is 
the possibility to perform higher-order computation with constant overhead. In particular, the universal function is realized
by a term $M_\mathit{eval}$ such
that $\app{M_\mathit{eval}}{\langle V,U\rangle}=\app{V}{U}$ and
$\timeapp{M_\mathit{eval}}{\langle V,U\rangle}=4+\timeapp{U}{V}$.
\condinc{
\begin{proposition}[Higher-Order]
There are terms $M_\mathit{eval},M_\mathit{curry}\in\Xi$ and constants 
$c_\mathit{eval}$, $c_\mathit{curry}^1$, $c_\mathit{curry}^2$, $c_\mathit{curry}^3$ 
such that, for every $V,U\in\Xi$, there
are $W,X,Y,Z\in\Xi$ such that:
\begin{eqnarray*}
\app{M_\mathit{eval}}{\langle V,U\rangle}&=&\app{V}{U}\\
\app{M_\mathit{curry}}{V}&=&W\\
\app{W}{X}&=&Y\\
\app{Y}{Z}&=&\app{V}{\langle X,Z\rangle}\\
\timeapp{M_\mathit{eval}}{\langle V,U\rangle}&\leq&c_\mathit{eval}+\timeapp{U}{V}\\
\timeapp{M_\mathit{curry}}{V}&\leq& c_\mathit{curry}^1\\
\timeapp{W}{X}&\leq& c_\mathit{curry}^2\\
\timeapp{Y}{Z}&\leq& c_\mathit{curry}^3+\timeapp{V}{\langle X,Z\rangle}
\end{eqnarray*}
\end{proposition}
\begin{proof}
Define:
\begin{eqnarray*}
  M_\mathit{eval}&\equiv&\lambda x.x(\lambda y.\lambda w.yw)\\
  M_\mathit{curry}&\equiv&\lambda x.\lambda y.\lambda w.x(\lambda z.zyw)\\
\end{eqnarray*}
Now, observe that
\begin{eqnarray*}
  M_\mathit{eval}\langle V,U\rangle &\timearrow{(1)}&\langle V,U\rangle(\lambda y.\lambda w.yw) \\
  &\timearrow{(1)}&(\lambda y.\lambda w.yw)VU\\
  &\timearrow{(1)}&(\lambda w.Vw)U\timearrow{(1)}VU\\
  M_\mathit{curry}V&\timearrow{(1)}&\lambda y.\lambda w.V(\lambda z.zyw)\equiv W\\
  WX&\timearrow{(1)}&\lambda w.V(\lambda z.zXw)\equiv W\equiv Y\\
  YZ&\timearrow{(1)}&V(\lambda z.zXZ)\equiv V\langle X,Z\rangle
\end{eqnarray*}
\end{proof}}{}
The fact that a ``universal'' combinator with a constant cost can be defined is quite
remarkable. It is a consequence of the inherent higher-order of the
lambda-calculus. Indeed, this property does not hold in the context of
Turing machines.
\section{Conclusions}
We have introduced and studied the  difference cost model for the pure, untyped,
 call-by-value lambda-calculus. The difference cost model satisfies the invariance
thesis, at least in its weak version~\cite{vanEmdeBoas90}.
We have given sharp complexity bounds on the simulations establishing
the invariance and giving evidence that the difference cost model is a parsimonious one. 
We do not claim this model is the definite word
on the subject. More work should be done, especially on 
lambda-calculi based on other evaluation models.

The availability of this cost model allows to reason on the complexity of call-by-value 
reduction by arguing on the structure of lambda-terms, instead of using complicated 
arguments on the details of some implementation mechanism. In this way, we could obtain 
results for eager functional programs without having to resort
to, e.g., a SECD machine implementation. 

We have not treated space. Indeed, the very definition
of space complexity for lambda-calculus---at least in a
less crude way than just ``the maximum ink used~\cite{Lawall96icfp}''---is 
an elusive subject which deserves better and deeper study.
 
\bibliographystyle{plain}

\end{document}